\documentclass{aastex}          %% The manuscript based on AASTeX v5.x
\usepackage{spr-astr-addons}    %% mimicing ASTR journal style
\usepackage{epsfig}
\usepackage{graphicx}
\usepackage{color}

\begin{document}

\title{The period ratio $P_{1}/P_{2}$ of torsional Alfv\'en waves with steady flows in spicules}

%% Running heads
\shorttitle{Torsional Alfv\'en waves}
\shortauthors{Ebadi et al.}

\author{H.~Ebadi}
\affil{Astrophysics Department, Physics Faculty,
University of Tabriz, Tabriz, Iran\\
e-mail: \textcolor{blue}{hosseinebadi@tabrizu.ac.ir}}
\and
\author{S.~Shahmorad}
\affil{Applied Mathematics Department, Mathematics Faculty,
University of Tabriz, Tabriz, Iran\\
e-mail: \textcolor{blue}{shahmorad@tabrizu.ac.ir}}
\and
\author{S.~Vasheghani Farahani}
\affil{Department of Physics, Tafresh University, P.O. Box 39518-79611, Tafresh, Iran\\
e-mail: \textcolor{blue}{S.Vasheghanifarahani@tafreshu.ac.ir}}

\begin{abstract}
The aim here is to model the standing torsional oscillations in solar spicules in the presence of  density stratification, magnetic field expansion, and steady flows. By implementing cylindrical geometry, the eigenfrequencies, eigenfunctions, and the period ratio $P_{1}/P_{2}$ of these waves is obtained for finite plasma-$\beta$.
The shifts created by the steady flow justifies the divergence of the observed period ratio for the first and second periods from the number $2$.
\end{abstract}

%-------------------------------------------------

\section{Introduction}
     \label{S-Introduction}
The solar atmosphere is designed by beautiful magnetic structures e.g., loops, arcades, filaments, jets, spicules, etc. The evolution in the observational instruments and satellites has made the observation of these magnetic structures more and more possible over the years \citep{Golub2007,Pesnel2012}. But the concept that is of our interest is the oscillations guided by these magnetic structures. Due to the transverse structuring in the solar atmosphere caused by these plasma eruptions, magnetohydrodynamic (MHD) waves are bound to propagate along them \citep{Tom2008}.
In the present study the magnetic structures under consideration are the spicules. Spicules are one of the most observed phenomena in the solar atmosphere especially in the chromosphere. Although spicules look very similar to coronal jets, but they are usually smaller, cooler, and denser.
The chromospheric spicules observed at the solar limb are reported to have speeds around
$20-25$~km~s$^{-1}$ propagating out towards the solar corona \citep{Tem2009}. The diameter of spicules ranges from $400$~km to $1500$~km, and their lengths ranges from $5000$~km to $9000$~km. The typical lifetime of spicules is between $5$ and $15$ minutes, with densities between $3.5\times10^{16}$ and $2\times10^{17}$ m$^{-3}$. Their temperatures are estimated between $5000$ and $8000$ K \citep{bec68, ster2000}.
In the context of the present study, it is the wave dynamics in spicules which is to be taken under consideration. Spicules experience transverse oscillations which have already been observed by both, spectroscopic and imaging devices. Ground based coronagraphs \citep{nik67,Kukh2006,Tem2007} together with SOHO proved adequate for the detection of Doppler shift oscillations in spicules \citep{xia2005}. With the enhancement of observational devices (Hinode/SOT) periodic displacements of the spicule axis became more and more pronounced, where some where interpreted as kink \citep{Kim2008,he2009,Ebadi2014b,Ebadi2012a,Ebadi2013} and some as Alfv\'{e}nic waves \citep{De2007,Ebadi2014a}. However, the statistical study carried out by \citet{Okamoto2011} using Hinode/SOT showed that $59\%$ percent of the waves guided by spicules propagate upward, $21\%$ of the waves propagate downward, and $20\%$ of the waves are standing oscillations.
In the present study the MHD wave under consideration is the torsional Alfv\'en wave. Out of all the MHD waves \citep{Edwin1983}, only the torsional Alfv\'en wave is completely incompressible \citep{Van2008}. However \citet{Vash2010} showed that in the presence of an equilibrium magnetic twist, the torsional wave becomes compressible. Observation wise, there is still vague indirect evidence of the detection of torsional Alfv\'en waves in the solar chromosphere and corona. This is due to lack of spatial resolution in solar observations. However, \citet{Tem2003,Tem2007} studied the non-thermal broadening of the green line in a coronal loop and came to the conclusion that they are standing torsional waves. In the photosphere-chromosphere region, \citet{jes09} reported the periodic variations of the spectral line widths as detection of torsional waves.\\
One of the features of coronal seismology is the period ratio, $P_{1}/P_{2}$, of the waves, where $P_{1}$ represents its fundamental period and $P_{2}$ represents its first harmonic. In this line \citet{Ebadi2014a} analyzed the time series of oxygen line profiles obtained from SUMER/SOHO on the solar south limb spicules. They calculated the Doppler shifts and consequently Doppler velocities in the coronal hole region. Their wavelet analysis determined periods of the fundamental mode and
first harmonic. The calculated period ratios showed departures from the canonical value of $2$. The reason for this deviation could be due to two issues. One is density stratification \citep{Andries2009}, the other is magnetic twist \citep{Karami2009,Karami2012}. In other words, \citet{Van2007} analyzed active region observations based on TRACE instrument and detected two periods with high confidence and calculated $P_{1}/P_{2}$ as $1.8$.
Nonetheless, \citet{Ebadi2014c} studied the competitive effects of density stratification and magnetic field expansion on torsional Alfv\'en waves in solar spicules. They showed that under some circumstances this ratio can approach its observational value even though it departs from its canonical value of $2$.

In the present study, in addition to the density stratification and magnetic field expansion of the cylindrical structure which resembles a solar spicule, a steady flow parallel to the vertical magnetic field is also taken in to account. Prior to this work the effects of steady flows where studied in cylindrical structures by \citet{goos1992,Terra2003,vash2009} where it was shown how the steady flow makes the frequencies undergo a Doppler shift as well as shifting the cut-off frequencies. The consequence of this could be reflected in the period ratios of the oscillations in the wave guiding structure \citet{Ruderman2010}. \citet{Ebadi2012b} analyzed the observational data obtained from
Hinode/SOT and performed time slice diagrams shedding more light on the spicule oscillations.
In the present study the MHD waves in the presence of steady flow inside solar spicule is studied. The model is discussed in the next section.

\section{Model and equilibrium conditions}
Consider an expanding untwisted magnetic flux tube with varying density along its axis. If a steady flow along the axis of this tube is present, a solar spicule is resembled. It is now clear that the geometry implemented will be cylindrical with coordinates $(r,\varphi,z)$. Note that the cylinder axis is aligned with the $z$-axis and the equilibrium magnetic field components are $B_{r}$ and $B_{z}$ with no azimuthal field, $(B_{\varphi}=0)$. The equilibrium magnetic field, $B$, could be written in terms of the vector potential ($A$) as
\begin{equation}
\label{eq:mage}
\mathbf{A} = \frac{\psi(r,z)}{r} \hat{e}_{\varphi},
\end{equation}
where we have
\begin{eqnarray}
\label{eq:f2}
 B_{r}  = -\frac{1}{r} \frac{\partial \psi}{\partial z} ,\,\,\,\,\,\,\,\,\,\,\,\,\,\,\
 B_{z}  = \frac{1}{r} \frac{\partial \psi}{\partial r}.
\end{eqnarray}
Note that $\psi$ is taken constant along the field lines.
Since the aim here is to study standing torsional Alfv\'{e}n waves, we consider one node on the foot point of the spicule at $z=0$ and the other node on the extent of the spicule at $z=L$. Note that $L$ is the spicule length. The dynamics of such oscillations are well described by the model proposed by
\citet{Ruderman2011}. If the steady flow inside the spicule is parallel to the spicule axis and represented by $v_{0}$, we have

\begin{equation}
\label{eq:velo}
\left( \frac{\partial}{\partial t} + v_{0} \frac{\partial}{\partial z}\right )^{2} \mathbf{\xi} = \frac{1}{\mu_{0}\rho} (\nabla \times \mathbf{b})\times
\mathbf{B},
\end{equation}
and
\begin{equation}
\label{eq:mag}
\mathbf{ b} = \nabla \times(\mathbf{\xi} \times \mathbf{B}),
\end{equation}
where the plasma displacement is $\mathbf{\xi}=(0,\xi_{\varphi},0)$, and the magnetic field perturbation is $\mathbf{b}=(0,b_{\varphi},0)$. By considering the perturbations proportional to $\textrm{exp}(-i\omega t)$, Eqs.~(\ref{eq:velo}) and ~(\ref{eq:mag}) would become

\begin{eqnarray}
\label{eq:velo1}
 \mu_{0} \rho \left( \omega^{2} + 2i\omega v_{0} \frac{\partial}{\partial z} - v_{0}^{2} \frac{\partial^{2}}{\partial z^{2}} \right ) \xi_{\varphi} \nonumber\\
  +  \frac{B_{r}}{r} \frac{\partial (rb_{\varphi})}{\partial r} + B_{z} \frac{\partial b_{\varphi}}{\partial z} = 0,
\end{eqnarray}
and
\begin{equation}
\label{eq:mag1}
b_{\varphi} = \frac{\partial (B_{r}\xi_{\varphi})}{\partial r} + \frac{\partial (B_{z}\xi_{\varphi})}{\partial z}.
\end{equation}
In the long-wavelength limit where, $R$, the radius of expanding spicule is much smaller than the spicule length $(R/L\ll 1)$, Eqs. (\ref{eq:velo1}) and (\ref{eq:mag1}) could be combined to give

\begin{eqnarray}
\label{eq:full}
  \mu_{0} \rho \left( \omega^{2} + 2i\omega v_{0} \frac{\partial}{\partial z} - v_{0}^{2} \frac{\partial^{2}}{\partial z^{2}} \right ) \xi_{\varphi}  &+&  \nonumber\\
  \frac{B_{z}}{R}\frac{\partial}{\partial z}
 \left[R^{2} B_{z} \frac{\partial}{\partial z} \left(\frac{\xi_{\varphi}}{R}\right)\right] &=& 0.
\end{eqnarray}
In obtaining Equation (\ref{eq:full}) the independent variable $\psi$ has been used instead of $r=r(\psi,z)$ by implementing the following rules
\begin{eqnarray}
\label{eq:f1}
  \frac{\partial f}{\partial r} &=& rB_{z} \frac{\partial f}{\partial \psi} \nonumber\\
  \frac{\partial f}{\partial z} &=& \frac{\partial f}{\partial z} - rB_{r} \frac{\partial f}{\partial \psi}.
\end{eqnarray}
Now by differentiating the identities $\psi=\psi(r(\psi,z),z)$ and $r=r(\psi(r,z),z)$ with respect to $z$, and making use of Equation (\ref{eq:f2}), one obtains
\begin{eqnarray}
\label{eq:f3}
  \frac{\partial r}{\partial z} &=& \frac{B_{r}}{B_{z}} \nonumber\\
  \frac{\partial r}{\partial \psi} &=& \frac{1}{rB_{z}},
\end{eqnarray}
see \citet{Ruderman2008,Verth2010,Karami2011} for details. Equation~(\ref{eq:full}) is a generalization of the analysis presented by \citet{Morton2011}.

In the context of coronal seismology, it would be more realistic to use the background magnetic field, plasma density, and spicule radius inferred from the actual magnetoseismology of observation \citep{Verth2011}

\begin{eqnarray}
\label{eq:equlib}
  B_{z}(z) &=& B_{0}\exp (-z/H_{B}) \nonumber\\
  \rho(z) &=& \rho_{0}\exp (-z/H_{\rho}) \nonumber\\
  R(z) &=& R_{0}\exp (z/2H_{B}).
\end{eqnarray}
The third expressio of Equation~\ref{eq:equlib} is due to the conservation of the magnetic flux, where
$H_{B}$ and $H_{\rho}$ are the magnetic and density scale heights, respectively.
substituting Eqs.~(\ref{eq:equlib}) in Equation~(\ref{eq:full}) gives

\begin{eqnarray}
\label{eq:pde}
  \left(1-M_{A}^{2} e^{-\alpha z} \right) \frac{\partial^{2} \xi_{\varphi}}{\partial z^{2}}  &+&  \nonumber\\
  \left(4\pi i M_{A} e^{-\alpha z} - \frac{1}{H_{B}} \right) \frac{\partial \xi_{\varphi}}{\partial z} &+&  \nonumber\\
  \left( \frac{1}{4H^{2}_{B}} + 4 \pi ^{2} \omega ^{2} e^{-\alpha z}\right)\xi_{\varphi} &=& 0,
\end{eqnarray}

where $\alpha\equiv\left(\frac{H_{B} - 2H_{\rho}}{H_{\rho} H_{B}}\right)$ and $M_{A} \equiv \frac{v_{0}}{v_{A}}$.
In Equation (\ref{eq:pde}) the lengths are normalized by the spicule length ($L$), and the frequencies by  the Alfv\'{e}n frequency
($\omega_{A}\equiv\frac{v_{A}}{L}=0.06$ rad/s). The other parameters are as; $v_{A}=\frac{B_{0}}{\sqrt{\mu_{0} \rho_{0}}}=75$ km/s; $B_{0}=12$ G,
$\rho_{0}=1.9\times10^{-10}$ kg m$^{-3}$, $\mu_{0}=4\pi\times10^{-7}$ T m A$^{-1}$, and $L=8000$ km).
The magnetic and density scale heights have been determined by \citet{Verth2011} as $H_{B}=0.1135$ and $H_{\rho}=0.094$, respectively.

\section{Numerical results and Discussions}

The numerical study carried out here is based on the differential transform method
(DTM \citep{Erturk2012,Nazari2010}). This method is applied to Equation~(\ref{eq:pde}) which gives the   eigenfrequencies and eigenfunctions of standing torsional Alfv\'{e}n
waves in stratified and expanding solar spicules, see \citet{Ebadi2014c} for detailed description of this method.

\begin{figure}
\centering
\includegraphics[width=8cm]{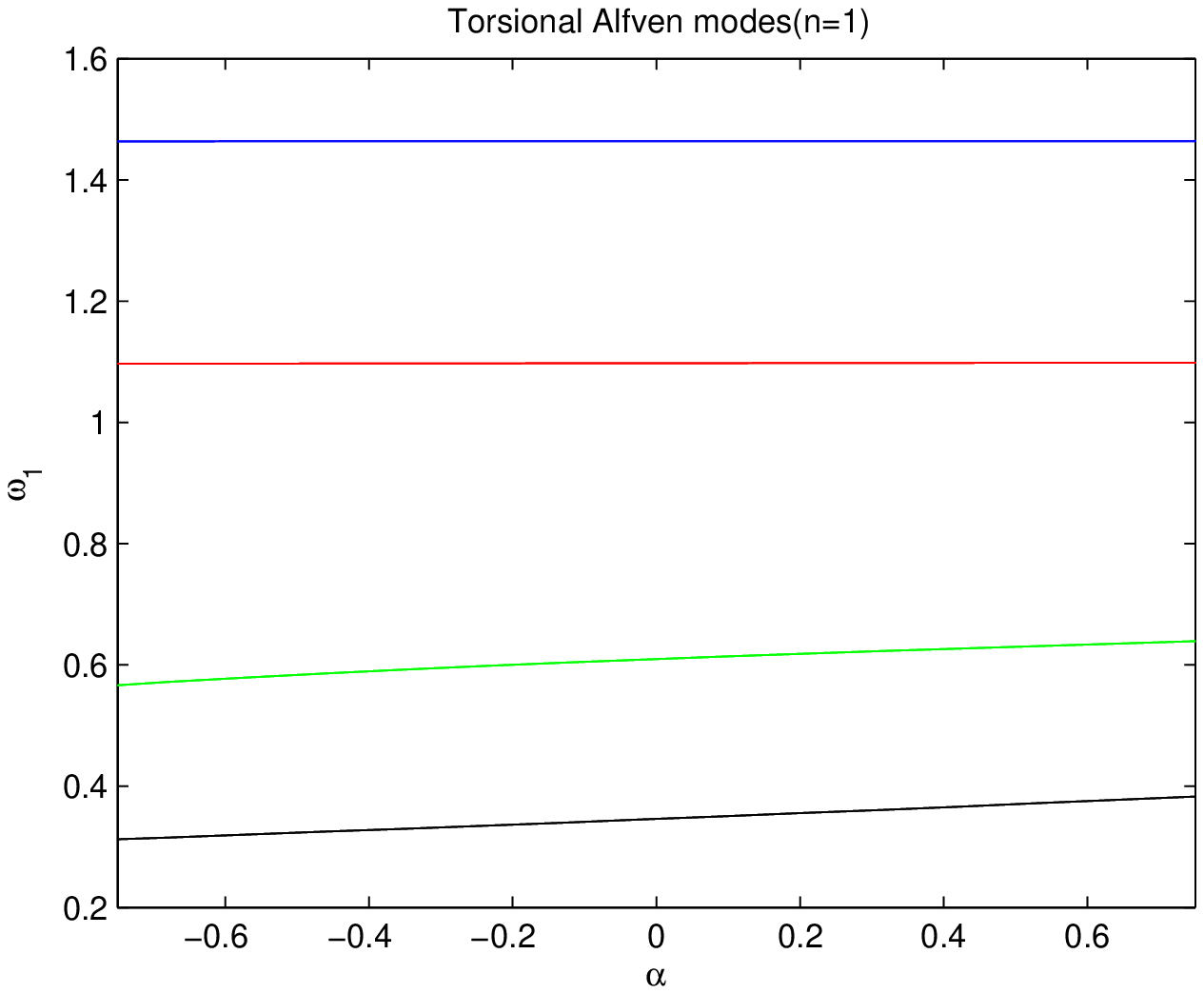}
\includegraphics[width=8cm]{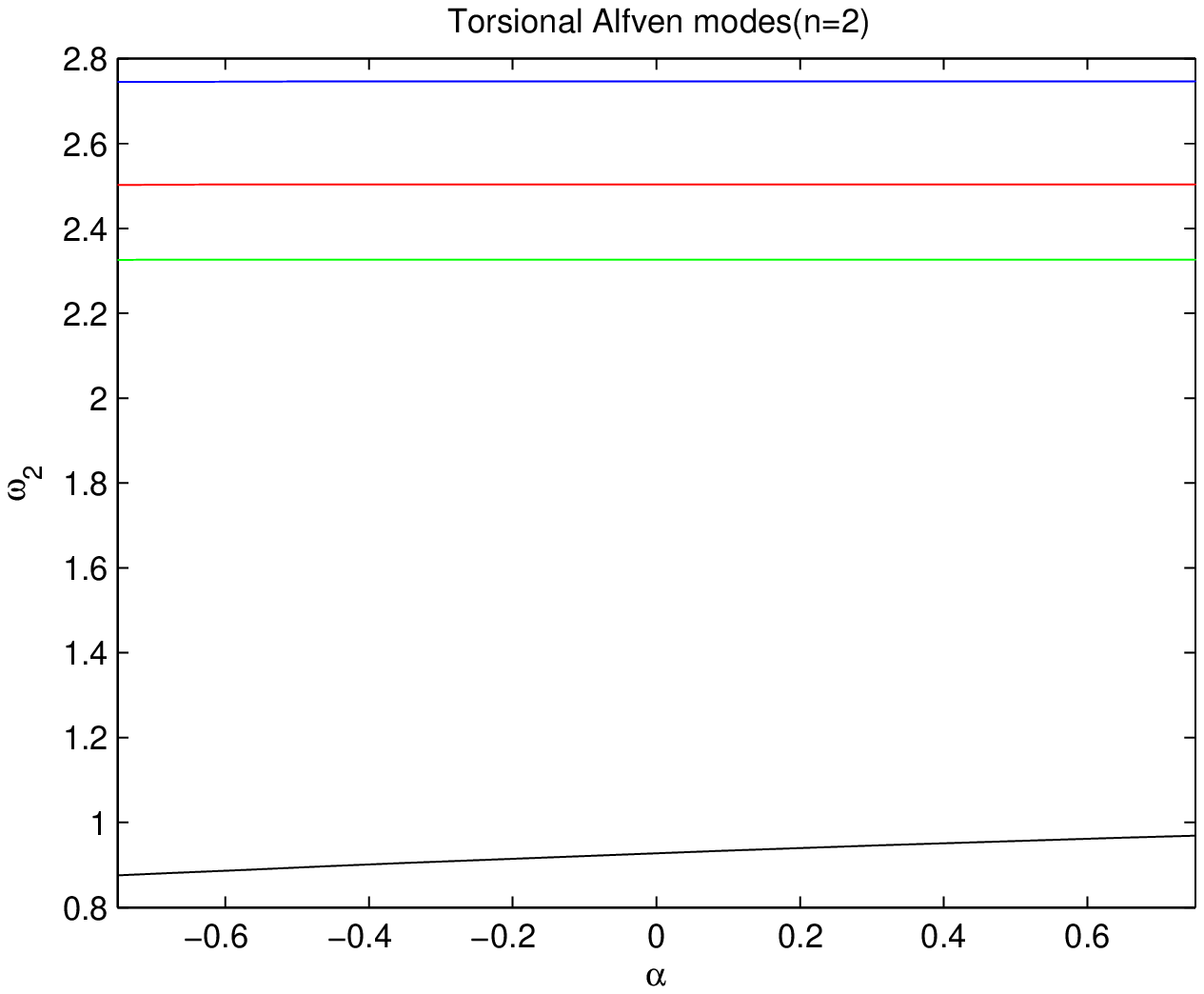}
\includegraphics[width=8cm]{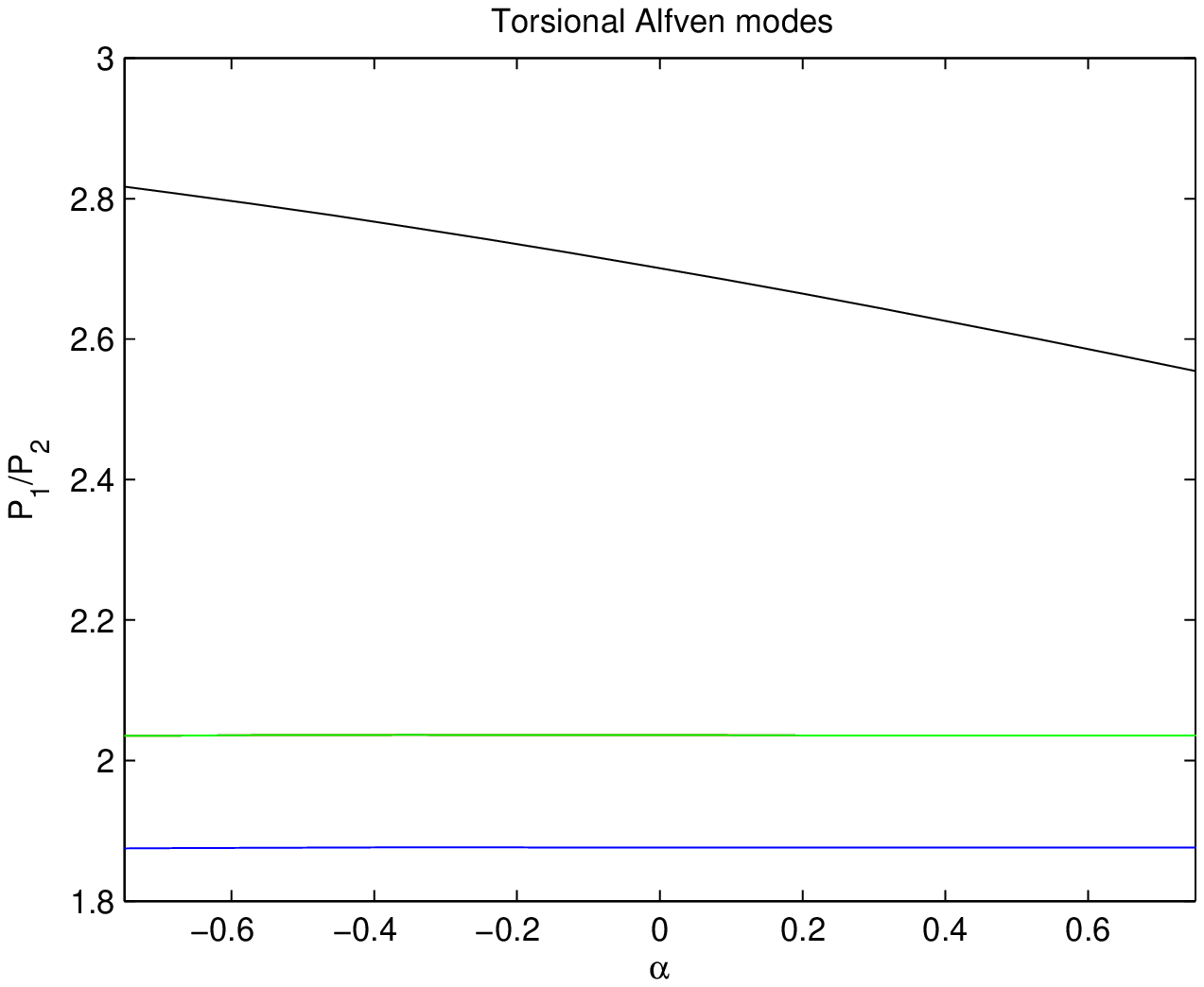}

\caption{Top panel, the fundamental torsional Alfv\'{e}n wave frequency for Mach numbers equal to $M_{A}=0, 0.2, 0.3, 0.4$ plotted against the parameter $\alpha$. Middle panel, the second harmonic frequency plotted against the parameter $\alpha$ . The bottom panel, the period ratio $P_{1}/P_{2}$ of
the fundamental $P_{1}$ and second harmonic period $P_{2}$ of the torsional wave against the parameter $\alpha$. Note that the colors black, blue, red, and green correspond to $M_{A}=0, 0.2, 0.3, 0.4$, respectively.  }
   \label{fig1}
   \end{figure}

We use the rigid boundary conditions and assume that $\xi_{\varphi}(0)=\xi_{\varphi}(L)=0$.
In the top panel of Figure~\ref{fig1} the fundamental frequency $(n=1)$ of the torsional Alfv\'{e}n wave is plotted with respect to the parameter $\alpha$ for various Mach numbers. This has been repeated for the second harmonic and shown in the middle panel of Figure~\ref{fig1}. It could be readily noticed by comparing the corresponding curves of the top and middle panels of Figure~\ref{fig1} that their ratios are away from number $2$. Now to give an exact flavor, the period ratios of the first and second harmonics, $P_{1}/P_{2}$, are plotted in the bottom panel of Figure~\ref{fig1}.

As it is expected, the frequencies are shifted not only because of the
flow, but also because of stratification and tube expansion \citep{Ebadi2012b}.
In other words, their variations respect to $\alpha$ are very moderate. 
Both the effects of stratification as well as the tube expansion are quite strong.
However they act in the opposite directions. While the stratification results in the
reduction of the ratio $P_{1}/P_{2}$, the tube expansion leads to the increase of this ratio.
As a result, the two effects almost cancel each other. We can observe that they
exactly cancel each other when $H_{B}=2H_{\rho}$.

It could be noticed that slower steady flows create higher frequencies and consequently lower periods with respect to those of static medium. An interesting result which is obtained by looking at the bottom panel of Figure~\ref{fig1} is that the period ratio, $P_{1}/P_{2}$, is very close to the observed values. This is due to the consideration of the steady flows in the analysis. As a matter of fact, in the case of Mach numbers equal to $0.2$ and $0.3$ the results are in very good agreement with observational results. It should be emphasized that steady flows have already been reported in observations carried out on spicules.

\begin{figure}
\centering
\includegraphics[width=8cm]{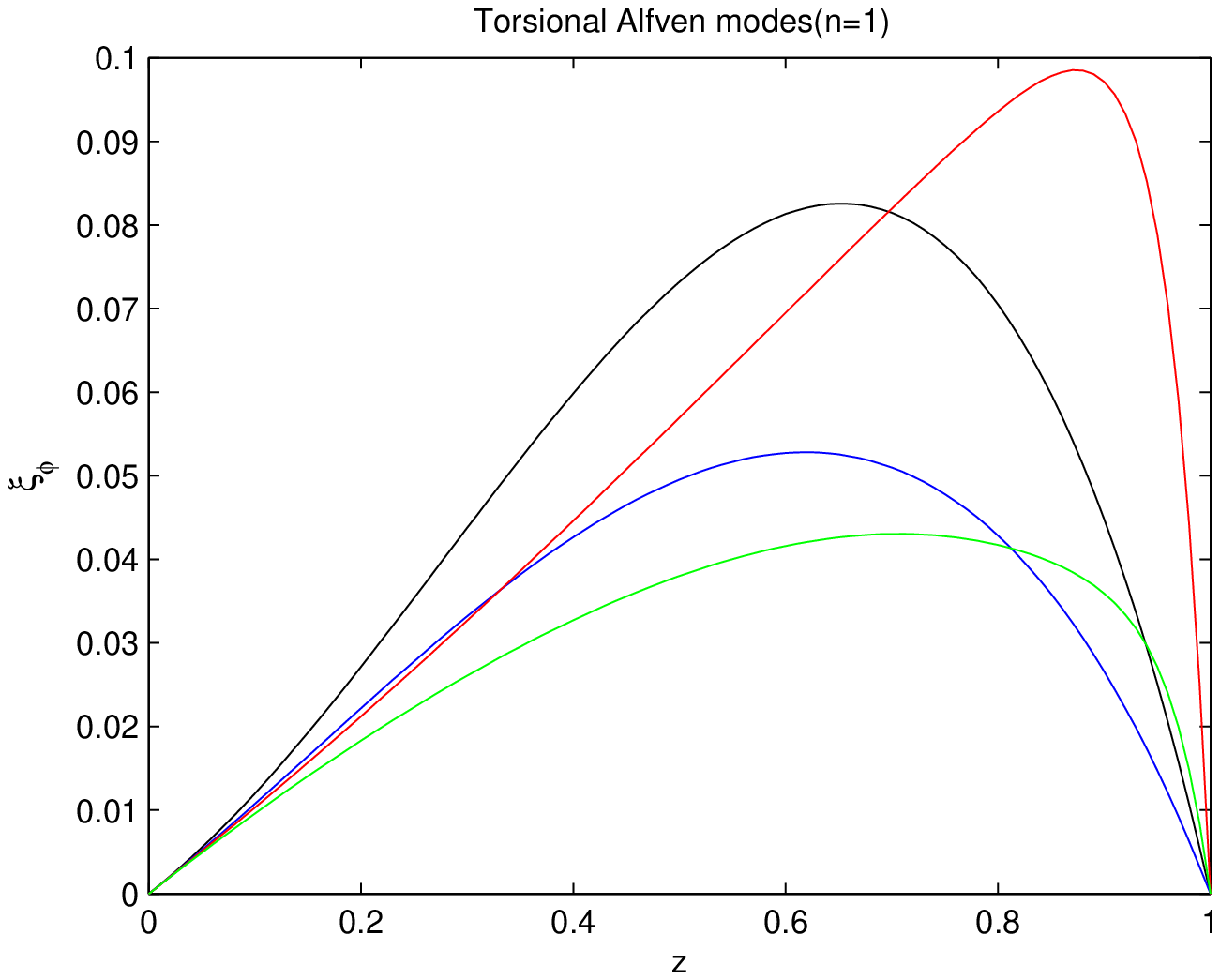}
\includegraphics[width=8cm]{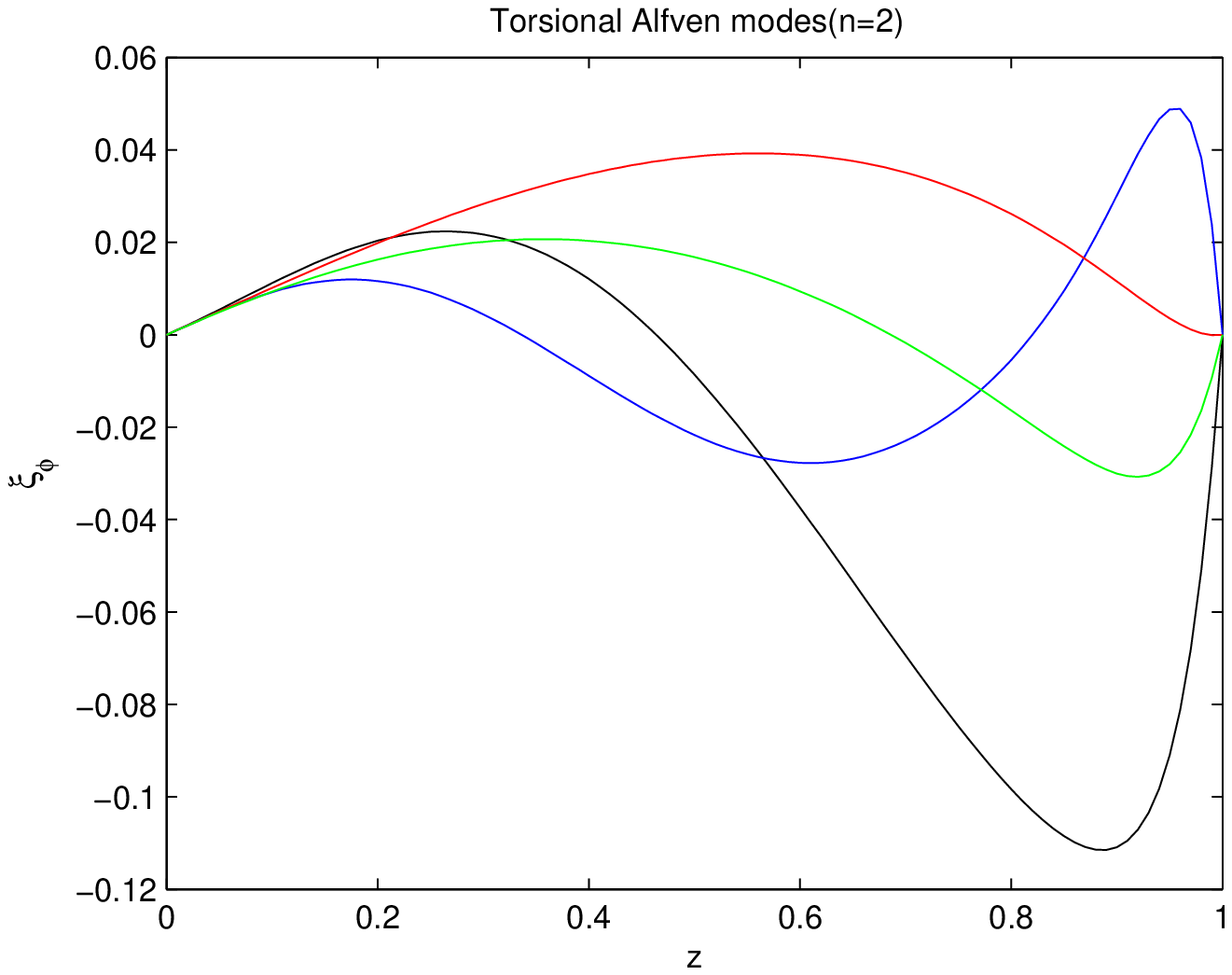}
\caption{Eigenfunctions of the fundamental and first harmonic torsional Alfv\'{e}n waves are plotted with respect to the normalized height along the spicule. The top panel corresponds to the fundamental oscillation where the bottom panel corresponds to the second harmonic. Note that the values of the parameter $\alpha$ and $M_{A}$ have been taken equal to $1.84$ and $0.25$, respectively. The order of colors are the same as of Figure~\ref{fig1}.
        }
   \label{fig2}
   \end{figure}

Eigenfunctions of the fundamental and first harmonic torsional Alfv\'{e}n modes with respect to the normalized height along the spicule
are presented in Figure~\ref{fig2}. We plotted eigenfunctions for $H_{B}=0.1135$, $H_{\rho}=0.094$, and $M_{A}=0.25$ which is determined by
observations.

\section{Conclusion}
      \label{S-Conclusion}
In the context of coronal seismology the aim is to understand the aspects of a phenomenon taking place in the corona by means of waves. The phenomenon under consideration in this study is the solar spicules and the wave for studying it is the torsional Alfv\'{e}n wave. The models that we implemented to study the spicule is an expanding untwisted magnetic flux tube with varying density along its axis with a steady flow. The coordinates implemented for modeling the spicule was cylindrical $(r,\varphi,z$).

 By implementing cylindrical geometry, the eigenfrequencies, eigenfunctions, and the period ratio $P_{1}/P_{2}$ of standing torsional waves was obtained based on the observational data provided by \citet{Verth2011} for finite plasma-$\beta$.
The shifts created by the steady flow justifies the divergence of the observed period ratio for the first and second periods from the number $2$. By plugging in the model, actual observed data of the  background magnetic field, plasma density, and spicule radius; the eigenfrequencies, eigenfunctions, and the period ratio $P_{1}/P_{2}$ of standing torsional waves were estimated. An interesting result  obtained is that the period ratio, $P_{1}/P_{2}$, is very close to the observed values. This is because of taking in to account the effects of steady flows in the analysis. As a matter of fact, in the case of Mach numbers equal to $0.2$ and $0.3$ the results get very close to observations. The interplay of stratification and equilibrium steady flow proved that the steady flow is something not to neglect. The good agreement of the numerical results obtained in this study with observations proof the efficiency of the steady flow effects compared to stratification.

\acknowledgments
This work is published as a part of the research project supported by the University of Tabriz research affairs office. 
Authors thank anonymous referee for his/her useful comments in improving the article.

\end{document}